# Large amplitude microwave emission and reduced nonlinear phase noise in Co$_2$Fe(Ge$_{0.5}$Ga$_{0.5}$) Heusler alloy based pseudo spin valve nanopillars

Jaivardhan Sinha, Masamitsu Hayashi[*], Yukiko K. Takahashi, Tomohiro Taniguchi, Maksim Drapeko, Seiji Mitani and Kazuhiro Hono

*National Institute for Materials Science, Tsukuba 305-0047, Japan*

We have studied microwave emission from a current-perpendicular-to-plane pseudo spin valve nanopillars with Heusler alloy Co$_2$Fe(Ga$_{0.5}$Ge$_{0.5}$) electrodes. Large emission amplitude exceeding 150 nV/Hz$^{0.5}$, partly owing to the large magnetoresistance, and narrow generation linewidth below 10 MHz are observed. We also find that the linewidth shows significant dependence on the applied field magnitude and its angle within the film plane. A minimum in the linewidth is observed when the slope of the frequency versus current becomes near zero. This agrees with theoretical prediction that takes into account non-linear phase noise as a source for linewidth broadening.

[*]E-mail address: hayashi.masamitsu@nims.go.jp

Since the prediction that spin transfer torque (STT)[1-2] can induce magnetization dynamics, numerous experimental and theoretical studies have been performed[3]. Spin torque oscillators (STO)[4-6], a nanoscale frequency tunable microwave source, can be applied to microwave sources for electro-communication devices and oscillating field sources for microwave assisted recording heads[7]. To obtain narrow spectral linewidth and large microwave power, various ferromagnetic materials and different geometries have been investigated. Geometry-wise, it has been demonstrated that it is possible to further increase the output power and achieve narrow linewidth when multiple STOs are synchronized[8-10]. From the materials perspective, it is generally known that giant magnetoresistance (GMR) based STOs can generate narrow spectral linewidth but limited output amplitude[4-6], whereas magnetic tunnel junction (MTJ) based STOs are able to produce large emission amplitude but with larger linewidth[11-13], although recent studies report significant progress on the linewidth of MTJs based STOs[12,14-15].Here we report implementation of Heusler alloy based STO in a current perpendicular to plane (CPP) giant magnetoresistance (GMR) pillar device and demonstrate its ability to generate significantly narrow linewidth along with large peak amplitude.

In general, Heusler alloys show large GMR[16], possess small magnetic damping[17], and expected to have large spin polarization[18] that can reduce the threshold current to cause magnetization precession. These characteristics are all favorable for STO applications. However, up to date, there is no report on STO that uses Heusler alloy as the magnetic free layer[19]. Recently, we have reported that a quaternary Heusler alloy $Co_2Fe(Ga_{0.5}Ge_{0.5})$ shows significantly large GMR ratio owing to its high bulk spin asymmetry (~0.77)[20]. Using this material, here we show that emission amplitude exceeding 150 $nV/Hz^{0.5}$ and generation linewidth below 10 MHz can be obtained CPP-GMR nanopillar devices. Partly owing to the relatively large GMR, the emission amplitude is nearly ten times larger than the reported values in similar systems with NiFe free layer[21-22]. Moreover, we find that the narrowest linewidth (and the largest amplitude) spectrum is observed when the current dependence of the generation frequency is nearly zero, implying that the non-linear phase noise[23-27] is reduced.

CPP-GMR pillars are made from a film stack of 10 Cr|100 Ag|10 $Co_2Fe(Ga_{0.5}Ge_{0.5})$|5

Ag|2 $Co_2Fe(Ga_{0.5}Ge_{0.5})$|5 Ag|8 Ru (units in nm) deposited on a single crystalline MgO (001) substrate using magnetron sputtering. The stack was post annealed at 500 °C to improve the chemical ordering of the Heusler alloy. Electron beam (e-beam) lithography and Ar ion milling are used to define the pillar. Milling is carried out till the thicker $Co_2Fe(Ga_{0.5}Ge_{0.5})$ is etched out. The shape of the pillar is set to squares or rectangles with dimensions of 130×90 $nm^2$, 140×140 $nm^2$ and 240×140 $nm^2$. The corners of the pillars are rounded due to the comparatively low electron beam dose at these regions during the e-beam writing process. An exemplary image of our pillar is shown in the inset of Fig. 1(a).

A ground-signal-ground (GSG) contact pads are patterned to allow high frequency signal detection from the pillar. Microwave probes (bandwidth: DC-18 GHz) are used to contact the device and the signal is fed to a spectrum analyzer (bandwidth: 9 kHz to 20 GHz) via a bias tee and an amplifier (bandwidth: 0.1-15 GHz). The gain of the amplifier is assumed constant throughout the entire frequency range and the measured amplifier gain (~50 dB) is divided out to obtain the power from the pillar device. Negative current corresponds to flow of electron from the thin to thick $Co_2Fe(Ga_{0.5}Ge_{0.5})$ layer. The sample is subjected to in plane magnetic field which is generated by a two axis vector magnet. We define 0° of the field angle as the magnetic field pointing towards the +X direction in the inset of Fig. 1(a) and the sense of rotation is set to counter clockwise. We performed microwave emission measurements for a number of devices and here we show results from two representative devices of size 130×90 $nm^2$ (device A) and 140×140 $nm^2$ (device B).

To study the crystalline anisotropy of the film, in Fig. 1(b), we show the magnetization hysteresis loops of a continuous film with the same film stack structure on which the nanopillar devices were fabricated. Magnetization measurements are carried out using a vibrating sample magnetometry (VSM) and the applied field angle is varied within the film plane. From the hysteresis curves we infer that the film is isotropic, i.e. the shape of the magnetization hysteresis loop remains nearly unchanged at different applied field angle. Thus in the patterned pillar devices, the magnetic anisotropy within the film plane is dominated by the pillar shape. The saturation magnetizations for both thick and thin layers are estimated to be ~800-900 $emu/cm^3$ from separate measurements (data not shown).

Figure 1(a) shows the resistance as a function of applied field for device A. The field is applied along the X direction and the sense current is set to 10 µA, small enough to avoid any current related effects. For sweeping the field from large positive to negative values, the thin layer magnetization reverses and the pillar forms the antiparallel state at a field value of ~150 Oe, which more or less represents the magnitude of the dipole exchange coupling field between the two magnetic layers. Further increase in the field towards the –X direction results in the reversal of the thick layer magnetization at a field value of ~-315 Oe.

For the same device, the resistance as a function of current is shown in Fig. 1(c). The magnitude of the applied field is close to zero. In agreement with the convention of spin transfer torque, positive current stabilizes the parallel state whereas negative current favors the antiparallel state. We observe nearly a full switching from the parallel to antiparallel state at a current of $I_C^-$ ~0.2 mA and a rather gradual switching from the antiparallel to parallel state at $I_C^+$ ~2 mA. The mean switching currents, defined by $(I_C^+ + I_C^-)/2$, are plotted Fig. 1(d) as a function of the applied field along the X direction. The threshold current shows weak dependence on the applied field. The mean threshold current density $J_C$ in the field range shown in Fig. 1(d) is ~5.8x10$^6$ A/cm$^2$, which is comparatively small in metallic CPP-GMR nanopillars devices. Similar order of magnitude current densities were also observed in a different Heusler alloy based CPP-GMR device[28].

In Fig. 2, we show the spectral properties of device A. We first perform a coarse measurement in which we scan a wide range of frequency (0.1 to 20 GHz) to look for signals of magnetization precession when various combinations of in-plane magnetic field and current are applied. Note that in the coarse scan mode, the peak amplitude can be underestimated if the peak width is narrow since the number of points in the frequency axis is limited in the spectrum analyzer. Background environmental noise is subtracted by subtracting a spectrum measured with the same applied field and angle but with zero applied current. Overall, we find that negative current, which favors the anti-parallel alignment, can induce magnetization precession and emit microwave in the frequency range of ~10 GHz when relatively large magnitude of fields (more than ~100 Oe) are applied.

In Fig. 2(a), we show the peak amplitude as a function of applied field angle with fixed field magnitudes of 235 Oe, 295 Oe and 360 Oe. Note that when the field angle is varied, both the thin and thick layers respond to the field since the thick layer is not pinned. Thus the set angle does not represent the relative angle between the two layers. For clarity, we refer to the set angle as the "angle" hereafter. In Fig. 2(a), the current is set to -8 mA and the amplitude of the largest peak is plotted. We observe large microwave amplitude in the field angle range of ~90° to ~150° for a certain field magnitude. This angle range is not dependent on the field application history, i.e. the initial state of the magnetic configuration has little effect on the emission amplitude. Note that we do not observe any large amplitude peaks when the field direction is reversed. This type of asymmetry of the emission amplitude with respect to the field direction has also been reported in Ref. [29]. Details of the pillar shape as well as the inhomogeneity of the magnetic anisotropy can be the cause of such asymmetry.

To study the characteristics of the large amplitude emission spectra, the spectra are captured at a finer frequency resolution around the peak frequency (more than 10 times finer frequency step than the coarse one). In addition, to study the distribution of the precession mode, ten independent spectrum measurements are made for each value of current and field. Spectral parameters are extracted by fitting the peak with a Lorentzian function. We show the current dependence of the peak amplitude, frequency and linewidth in Fig. 2(b-d) for different applied field angles. Here the magnitude of the field is fixed to 235 Oe. In Fig. 2(b), the amplitude starts increasing above ~-5 mA and reaches a maximum before decreasing with further application of current. The maximum emission amplitude obtained in device A is ~155 nV/Hz$^{0.5}$. An exemplary spectrum is shown in the inset of Fig. 2(d). The integrated power for this peak amounts to ~5 nW, which is nearly ten times larger than previously reported values in similar systems[21-22]. Note that signal losses due to transmission across the cable and the internal resistance (~7-10 $\Omega$) of the pillar are not taken into account in the emission amplitude. With these corrections included, the amplitude will increase by a factor of ~1.1 to 1.2.

As shown in Fig. 2(c), the generation frequency slightly changes with increasing current in the range where the magnetization precession is observed. In the angle range shown, we find that the slope of the generation frequency versus current changes from "blue" (positive slope) to "red" (negative slope) as the current is increased, albeit the change is rather small. The linewidth, shown in Fig. 2(d), correlates with the emission amplitude; that is, it takes a minimum when the peak amplitude takes its maximum at each value of the applied field angle. The narrowest linewidth obtained here is ~10 MHz with a quality factor that exceeds 1000.

Note that the current at which the linewidth takes its minimum shifts as the angle is varied. Theoretically, it has been predicted that linewidth broadening occurs when the generation frequency shows significant dependence on the precession amplitude (and thus current)[23-27]. Such non-linear phase noise is expected to decrease when optimal field magnitude, angle and current is selected. Our results agree with this prediction: the minimum in the linewidth more or less coincides with the current at which the slope of the generation frequency versus square of current becomes zero. These current values are shown in Fig. 2(d) by the arrows for each field angle.  For each field angle above a particular current the linewidth starts broadening. However, the change in the generation frequency with current is rather small above this current and thus non-linear phase noise is expected to be small in this regime. We believe that the increase in the linewidth in the high current regime is due to the onset of the instability of the coherent precession modes[30-33].

It is not clear what factors plays a dominant role in determining the field magnitude and angle at which the generation linewidth takes its minimum. Theoretically it has been predicted that when the field is directed along the magnetic hard axis with its magnitude being four times the effective anisotropy field, the non-linear phase noise vanishes[34]. In device A, the angle at which we observe the maximum emission amplitude does not coincide with the hard axis defined by the pillar shape. The field dependence of the thick layer further complicates the angular dependence. The condition at which large amplitude emission is observed seems to vary from device to device. As an example, we show the spectral characteristics of device B in Fig. 3.

In Fig. 3(a) we show the angular dependence of the emission amplitude at an applied field of 435 Oe for a set current of -8 mA and 660 Oe for a set current of -9 mA. We observe large amplitude emission for a field angle of around ~80° and ~260°, and an additional peak at ~210° for higher fields. In Fig. 3(b-d), we show the current dependence of the emission amplitude, peak frequency and the linewidth for different magnetic field angles at a field magnitude of 660 Oe. The current dependence of the amplitude, frequency and linewidth show similar features as observed in device A. The generation frequency shows small change with current and there is a current at which the linewidth and the amplitude take their extrema. Moreover, in device B, we observe the generation frequency changes its slope against current when the angle is varied. The minimum linewidth (~4.9 MHz, Q-factor exceeding ~2,300) and the maximum emission amplitude take place at an angle where the slope becomes nearly zero. Such angular dependence of the linewidth agrees with theoretical predictions[34] and has been reported previously in other systems[15,22].

The minimum linewidth obtained here in our CPP-GMR pseudo spin valves with Heusler alloy $Co_2Fe(Ga_{0.5}Ge_{0.5})$ electrodes is 4-5 times smaller than that reported for conventional ferromagnetic electrodes when the non-linear phase noise is reduced[22]. This could be partly due to the precessing free layer accommodating larger spin torque without entering chaotic magnetization dynamics. Since the spin polarization of the $Co_2Fe(Ge_{0.5}Ga_{0.5})$ Heusler alloy is high (large bulk spin asymmetry $\beta$~0.77) compared to the conventional transition metal alloys[18], larger amount of spin torque can be applied to the free layer for a given current density. This manifests itself as the small threshold current density for magnetization switching, as shown in Fig. 1(d). In general, there is a threshold spin torque at which the coherence of the magnetization precession mode is disrupted and eventually enters chaotic modes. The exchange stiffness[35] of the $Co_2Fe(Ga_{0.5}Ge_{0.5})$ electrode is comparatively large (Curie temperature of the film is ~1080 K) and perhaps more importantly it is uniform within the layer, which may help in sustaining coherent precession with large spin torque. Controlling the onset of chaotic magnetization dynamics is important for developing high quality spin torque oscillators and further study is required to gain more understanding on this topic[36-37].

In summary, we have studied microwave emission from a current perpendicular to plane pseudo spin valve nanopillar with Heusler alloy $Co_2Fe(Ga_{0.5}Ge_{0.5})$ electrodes. By setting the magnetic field magnitude/angle and current to optimum value, we observe large emission amplitude exceeding 150 nV/Hz$^{0.5}$ and narrow generation linewidth below 10 MHz. The narrow linewidth (~4.9 MHz with quality factor exceeding 2,300) is obtained by reducing the non-linear phase noise. In addition, owing to the low threshold current density required to cause magnetization reversal, of the order ~6x10$^6$ A/cm$^2$, larger amount of spin torque is applied to the free layer which can result in narrower linewidth and large emission amplitude. This in turn indicates that the $Co_2Fe(Ga_{0.5}Ge_{0.5})$ electrode can tolerate larger amount of spin torque without breaking the coherence of the precession mode, perhaps due to the homogeneity of the exchange stiffness across the film. At higher currents, the linewidth broadens even though the non-linear phase noise is small. Identifying the precession modes at higher currents and understanding the onset of decoherence is required for further development of high quality spin torque oscillators.

**Acknowledgments**

We are very grateful to C. Kline for technical support on sample fabrication. This work was partly supported by Grant-in-Aids for Scientific Research (A) (No. 22246091), the JSPS through its "FIRST program" and PRESTO-JST.

**Figure captions**

Fig. 1 (a) Resistance as a function of field for device A. The field is applied along the X axis. Inset: scanning electron microscopy image of a $140{\times}140$ nm$^2$ pillar (device B). Definitions of X and Y axis are included. (b) Magnetization hysteresis loops of a continuous film at different in-plane field angles. (c) Resistance versus current for pillar A. A small field of ~11 Oe is applied along the X direction. (d) Variation of the switching current density for device A plotted as a function of field along the X direction.

Fig. 2 Spectral characteristics of device A. (a) Angular variation of the emission amplitude with current of -8 mA. (b) Emission amplitude, (c) peak frequency and (d) linewidth versus current at different field angles. The field magnitude is fixed to 235 Oe. The arrows indicate currents at which the non-linear phase noise vanishes for different field angles. Inset to (d): One of the highest emission amplitude spectrum. Line is fit with a Lorentzian function.

Fig. 3 Spectral characteristics of device B. (a) Angular variation of the emission amplitude with current of -8 mA for 435 Oe and -9 mA for 660 Oe. (b) Emission amplitude, (c) peak frequency and (d) linewidth as a function of current at different field angles. The magnitude of magnetic field is fixed to 660 Oe.

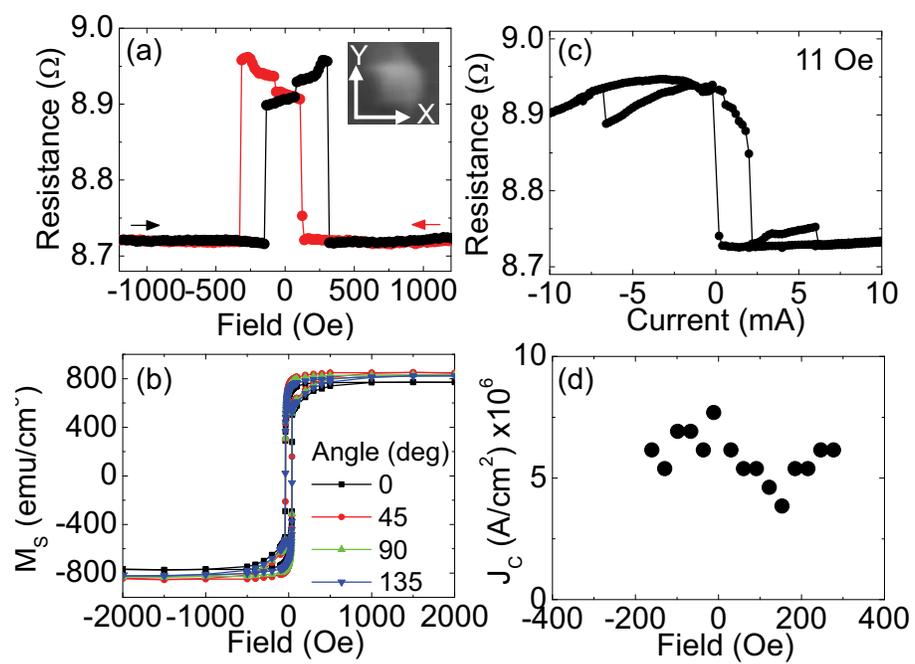

Fig. 1

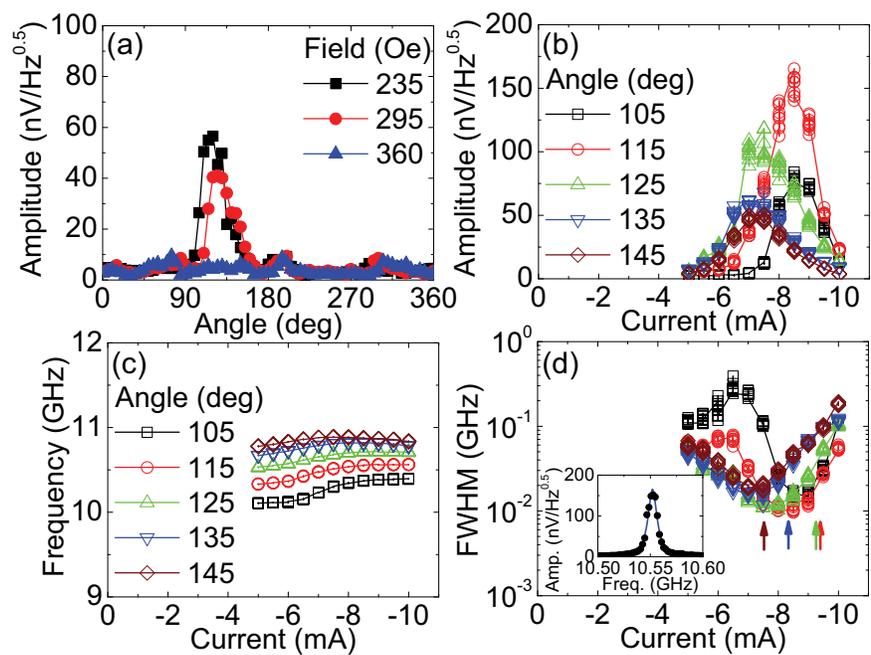

Fig. 2

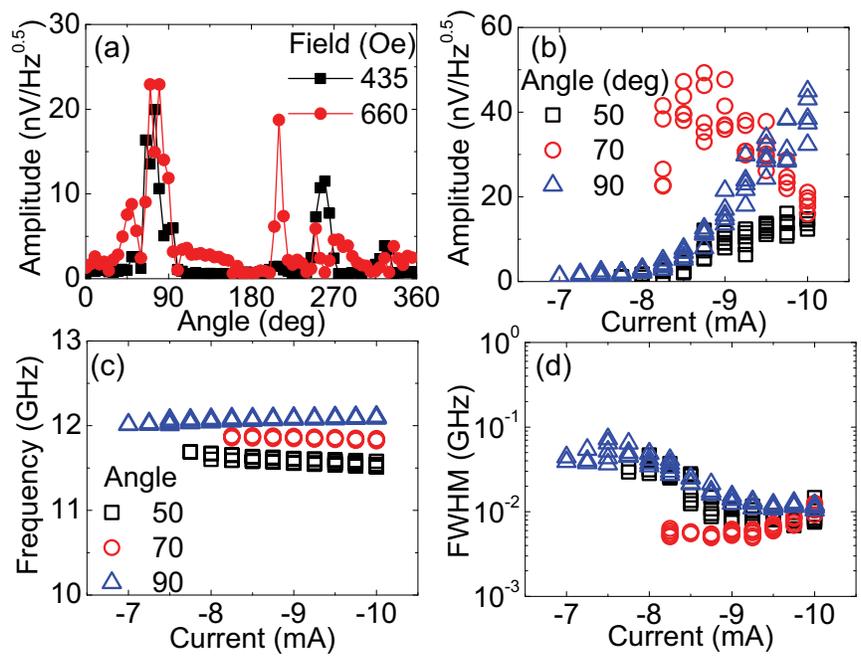

Fig. 3